# Maximum Willis Coupling in Acoustic Scatterers

Li Quan, Younes Ra'di, Dimitrios Sounas and Andrea Alù[*]

*Department of Electrical and Computer Engineering, The University of Texas at Austin,*

*Austin, Texas 78712, USA*

*Willis coupling in acoustic materials defines the cross-coupling between strain and velocity, analogous to bianisotropic phenomena in electromagnetics. While these effects have been garnering significant attention in recent years, to date their effects have been considered mostly perturbative. Here, we derive general bounds on the Willis response of acoustic scatterers, show that they can become dominant in suitably designed scatterers, and outline a systematic venue for the realistic implementation of maximally bianisotropic inclusions. We then employ these inclusions to realize acoustic metasurfaces for sound bending with unitary efficiency.*

The emergence of metamaterials and metasurfaces has enabled new opportunities to manipulate electromagnetic waves, providing a rich platform for extreme light-matter interactions [1]-[5]. The fascinating developments in this field have stimulated efforts to apply the same principles to waves of different physical nature, resulting in acoustic and elastic metamaterials. Owing to their unique properties, acoustic metamaterials have enabled unprecedented sound-matter interactions, such as cloaks of inaudibility, acoustic superlenses, acoustic collimators, and nonlinear acoustic phenomena [6]-[14]. In order to realize these artificial materials, various subwavelength resonant inclusions, such as

---

[*] Corresponding author: alu@mail.utexas.edu

Helmholtz resonators [7],[15], space-coiling structures [16],[17] and membrane-type inclusions [18], have been explored, to provide the required enhanced dipole and/or monopole responses.

In electromagnetic metamaterials, an additional knob to tailor the overall metamaterial response has been provided by magneto-electric coupling, or bianisotropy. In direct analogy, Willis coupling has been explored in elastodynamics [19]-[22] and acoustics [23]-[25] since the 80s, and it has been recently become of interest in the context of acoustic metamaterials. Willis coupling describes the connection between acoustic pressure and particle velocity in some acoustic media, and so far it has been treated as a higher-order perturbative phenomenon [22]-[24]. In order to exploit to its full extent bianisotropy in elastodynamics and acoustics, it would be ideal to realize metamaterial inclusions with large, ideally maximum, Willis coupling.

In this Letter, we analyze passive bianisotropic acoustic scatterers, defining a polarizability tensor that relates, in the most general linear scenario, monopole and dipole moments to the local pressure and velocity. We derive general bounds on the Willis coupling of small acoustic scatterers imposed by passivity and reciprocity, with direct implication on the maximum acoustic bianisotropy achievable in metamaterials formed by such inclusions. Our bound shows that the pressure-velocity coupling can, in principle, be of the same order as the direct polarizability terms, and we introduce a general framework to design subwavelength inclusions providing maximum Willis coupling. Finally, we apply these optimal inclusions to design bianisotropic metasurfaces, of great relevance to the problem of arbitrary wavefront transformation with large efficiency.

Consider a subwavelength particle located in a fluid background and excited by an acoustic wave. Since the particle is small, its scattering can be described by the superposition of acoustic monopole and dipole moments. For non-bianisotropic linear inclusions, the monopole is proportional to the local pressure field, while the dipole is proportional to the local velocity. In the case of Willis coupling, both pressure and velocity fields can excite monopole and dipole moments, and their general relation can be written as

$$\begin{pmatrix} M \\ \mathbf{D} \end{pmatrix} = \boldsymbol{\alpha} \begin{pmatrix} p \\ \mathbf{v} \end{pmatrix} = \begin{pmatrix} \alpha_{pp} & \boldsymbol{\alpha}_{pv} \\ \boldsymbol{\alpha}_{vp} & \boldsymbol{\alpha}_{vv} \end{pmatrix} \begin{pmatrix} p \\ \mathbf{v} \end{pmatrix}. \qquad (1)$$

Here $M = \int_V \rho dV$ is the acoustic monopole, $\mathbf{D} = \int_V \rho \mathbf{r} dV$ is the acoustic dipole moment, $p$ is the local pressure, $\mathbf{v}$ is the local velocity, $\boldsymbol{\alpha}$ is the polarizability tensor and $\rho$ is the density distribution in the particle. The off-diagonal terms $\boldsymbol{\alpha}_{pv}$ and $\boldsymbol{\alpha}_{vp}$ in the polarizability tensor, responsible for Willis coupling, arise when the particle is geometrically asymmetric, which couples pressure and velocity responses.

Based on these general relations, we can study fundamental constraints imposed by reciprocity and energy conservation over the polarizability tensor of an acoustic inclusion. For simplicity, and without loss of generality, we assume a two-dimensional (2D) scenario, for which the polarizabilies in Eq. (1) can be explicitly written as

$$\begin{pmatrix} M \\ D_x \\ D_y \end{pmatrix} = \begin{pmatrix} \alpha_{pp} & \alpha_x^{pv} & \alpha_y^{pv} \\ \alpha_x^{vp} & \alpha_{xx}^{vv} & \alpha_{xy}^{vv} \\ \alpha_y^{vp} & \alpha_{yx}^{vv} & \alpha_{yy}^{vv} \end{pmatrix} \begin{pmatrix} p \\ v_x \\ v_y \end{pmatrix}. \qquad (2)$$

We extend our results to 3D in [26]. Let us first examine the constraints on $\boldsymbol{\alpha}$ imposed by energy conservation. For passive scatterers, the total scattered power must be less or equal than the extinction power $\iint p_s \vec{v}_s^* \cdot d\vec{A} \leq -\iint \left( p_s \vec{v}_i^* + p_i \vec{v}_s^* \right) \cdot d\vec{A}$. For a subwavelength scatterer, monopole and dipole moments dominate the scattering, so this relation can be explicitly written as [26]

$$\omega^2 \left( \left|\sqrt{2}M\right|^2 + \left|ik_0 D_x\right|^2 + \left|ik_0 D_y\right|^2 \right) \leq 8 \operatorname{Im}\left[ \rho_0 c_0 \vec{v}^* \cdot \left(ik_0 \vec{D}\right) - \frac{p^*}{\sqrt{2}}\left(\sqrt{2}M\right) \right]. \tag{3}$$

Substituting the monopole and dipole moment expressions from Eq. (2), we get a general condition imposed by energy conservation over the polarizabilities of a general acoustic bianisotropic particle [26]

$$\operatorname{Diag}\left[ \omega^2 \left( \boldsymbol{\alpha}'^{*T} \boldsymbol{\alpha}' \right) \right] \leq \operatorname{Diag}\left[ 4i \left( \boldsymbol{\alpha}'^{T*} - \boldsymbol{\alpha}' \right) \right], \tag{4}$$

where $\boldsymbol{\alpha}'$ is the normalized polarizability tensor defined as

$$\boldsymbol{\alpha}' = \begin{pmatrix} \alpha_{pp}' & \alpha_x^{pv\prime} & \alpha_y^{pv\prime} \\ \alpha_x^{vp\prime} & \alpha_{xx}^{vv\prime} & \alpha_{xy}^{vv\prime} \\ \alpha_y^{vp\prime} & \alpha_{yx}^{vv\prime} & \alpha_{yy}^{vv\prime} \end{pmatrix} = \begin{pmatrix} -2\alpha^{pp} & -\sqrt{2}\alpha_x^{pv}/\rho_0 c_0 & -\sqrt{2}\alpha_y^{pv}/\rho_0 c_0 \\ ik_0 \sqrt{2}\alpha_x^{vp} & ik_0 \alpha_{xx}^{vv}/\rho_0 c_0 & ik_0 \alpha_{xy}^{vv}/\rho_0 c_0 \\ ik_0 \sqrt{2}\alpha_y^{vp} & ik_0 \alpha_{yx}^{vv}/\rho_0 c_0 & ik_0 \alpha_{yy}^{vv}/\rho_0 c_0 \end{pmatrix}. \tag{5}$$

This normalization ensures that all terms in the tensor have the same dimensions. For a reciprocal acoustic particle, the normalized polarizability tensor satisfies $\boldsymbol{\alpha}' = \boldsymbol{\alpha}'^{T-}$ [26].

In the non-bianisotropic limit, i.e., in the absence of off-diagonal terms in (5), the reciprocity condition is always satisfied, and Eq. (4) requires

$$\operatorname{Im}\left(1/\alpha^{pp\prime}\right) \leq -\omega^2/8 \text{ and } \operatorname{Im}\left(1/\alpha_{xx}^{vv\prime}\right) = \operatorname{Im}\left(1/\alpha_{yy}^{vv\prime}\right) \leq -\omega^2/8. \tag{6}$$

These constraints on the imaginary part of the polarizability physically correspond to radiation loss in the particle, and the inequalities become equalities when the particle has no absorption. In the general bianisotropic scenario, we can replace the full polarizability tensor into Eq. (4), resulting in the interesting relations

$$\begin{cases} \sqrt{\left|\alpha_x^{vp\prime}\right|^2 + \left|\alpha_y^{vp\prime}\right|^2} \leq \sqrt{\dfrac{-\dfrac{8}{\omega^2}\mathrm{Im}\left(1/\alpha^{pp\prime}\right)-1}{\left|1/\alpha^{pp\prime}\right|^2}} \leq \dfrac{4}{\omega^2} \\[2ex] \sqrt{\left|\alpha_x^{vp\prime}\right|^2 + \left|\alpha_{xy}^{vv\prime}\right|^2} \leq \sqrt{\dfrac{-\dfrac{8}{\omega^2}\mathrm{Im}\left(1/\alpha_{xx}^{vv\prime}\right)-1}{\left|1/\alpha_{xx}^{vv\prime}\right|^2}} \leq \dfrac{4}{\omega^2} \\[2ex] \sqrt{\left|\alpha_y^{vp\prime}\right|^2 + \left|\alpha_{xy}^{vv\prime}\right|^2} \leq \sqrt{\dfrac{-\dfrac{8}{\omega^2}\mathrm{Im}\left(1/\alpha_{yy}^{vv\prime}\right)-1}{\left|1/\alpha_{yy}^{vv\prime}\right|^2}} \leq \dfrac{4}{\omega^2} \end{cases} \quad . \quad (7)$$

Since $\left|\alpha_x^{vp\prime}\right|$ and $\left|\alpha_y^{vp\prime}\right|$ cannot be negative, from the first conditions we derive the same bound as (6) on the diagonal terms of the polarizability tensor, indicating that these conditions are general, and apply equally well to bianisotropic particles. $\mathrm{Im}\left(1/\alpha^{pp\prime}\right)$ and $\mathrm{Im}\left(1/\alpha^{vv\prime}\right)$ increase because of the Willis coupling according to Eq. (7), and in the lossless limit the equality signs in (6) do not hold because of bianisotropy.

The first condition in Eq. (7) imposes also a general bound on the bianisotropic coupling terms. If we assume that the particle is only resonant in the $y$ direction, so that $\left|\alpha_y^{vp\prime}\right| \gg \left|\alpha_x^{vp\prime}\right|$, this bound simplifies into $\left|\alpha_y^{vp\prime}\right| \leq 4\omega^{-2}$. Interestingly, Eq. (7) implies that this limit can be

reached only when the particle is at resonance, i.e., when $1/\alpha_{yy}^{vv'} = -i\omega^2/4$ and $1/\alpha^{pp'} = -i\omega^2/4$ are purely imaginary. Substituting these conditions back into the polarizability tensor, we find that optimal Willis scatterers, with maximum bianisotropic coupling, have the normalized polarizability tensor

$$\boldsymbol{\alpha}' = \frac{4}{\omega^2}\begin{pmatrix} i & 0 & e^{i\varphi} \\ 0 & 0 & 0 \\ -e^{i\varphi} & 0 & i \end{pmatrix}, \tag{8}$$

where $\varphi$ is an arbitrary phase factor. Interestingly, Eq. (8) confirms that Willis coupling can become of the same order as the diagonal elements of the polarizability tensor, opening new opportunities to enable strong bianisotropy in small acoustic scatterers and metamaterial inclusions. In [26], we extend this analysis and the associated bounds to 3-D inclusions, deriving the analogous bound $\left|\alpha_y^{vp'}\right| \leq \dfrac{6\pi}{c_0^2 k_0^3}$.

From Eq. (7), we can derive the general bound on the cross-coupling polarizability terms

$$\sqrt{\frac{\left|\alpha_x^{vp'}\right|^2 + \left|\alpha_y^{vp'}\right|^2}{\left|\alpha^{pp'}\right|\sqrt{\left|\alpha_{xx}^{vv'}\right|^2 + \left|\alpha_{yy}^{vv'}\right|^2}}} \leq$$

$$\sqrt[4]{\left(-\frac{8}{\omega^2}\operatorname{Im}\left(1/\alpha^{pp'}\right)-1\right)\left[\frac{\left(-\frac{8}{\omega^2}\operatorname{Im}\left(1/\alpha_{xx}^{vv'}\right)-1\right)\left|\alpha_{xx}^{vv'}\right|^2 + \left(-\frac{8}{\omega^2}\operatorname{Im}\left(1/\alpha_{yy}^{vv'}\right)-1\right)\left|\alpha_{yy}^{vv'}\right|^2 - 2\left|\alpha_{xy}^{vv'}\right|^2}{\left|\alpha_{xx}^{vv'}\right|^2 + \left|\alpha_{yy}^{vv'}\right|^2}\right]}$$

(9)

If we suppose again a dominant response along y, so that $\alpha_x^{vp'} \approx 0$, $\alpha_{xx}^{vv'} \approx 0$ and $\alpha_{xy}^{vv'} \approx 0$, this condition simplifies into

$$\frac{\left|\alpha_x^{vp'}\right|}{\sqrt{\left|\alpha^{pp'}\right|\left|\alpha_{yy}^{vv'}\right|}} \leq \sqrt[4]{\left(-\frac{8}{\omega^2}\mathrm{Im}\left(1/\alpha^{pp'}\right)-1\right)\left(-\frac{8}{\omega^2}\mathrm{Im}\left(1/\alpha_{yy}^{vv'}\right)-1\right)}, \qquad (10)$$

indicating that, while there is no limit to the normalized Willis response compared to the direct response of the scatterer to pressure and velocity, at resonance when $\left|\alpha_x^{vp'}\right|$ reaches its maximum value $4\omega^{-2}$, we get

$$\left|\alpha_x^{vp'}\right| \leq \sqrt{\left|\alpha^{pp'}\right|\left|\alpha_{yy}^{vv'}\right|}, \qquad (11)$$

As an excursus, for electromagnetic waves Sersic *et al.* derived a condition on the off-diagonal (bianisotropic) polarizability elements that cannot be greater than the square root of the product of the diagonal coupling coefficients [30]-[31], consistent with Eq. (11), under the assumption of a single resonance. However, recently Albooyeh *et al.* reported that some non-resonant structures can break these restrictions [32]. Our finding in (10), if translated to electromagnetic waves, addresses this issue showing that, at resonance, for large bianisotropy, Eq. (11) holds, but off resonance it is possible to yield larger relative off-diagonal elements, at the cost of sacrificing their overall magnitude. In addition, our derivation extends and generalizes these relations to acoustic problems.

Our general analysis so far shows that Willis coupling in a small resonant scatterer can be as strong and important as the diagonal terms of the polarizability tensor. To exploit the possibilities offered by bianisotropic cross-coupling and create unprecedented sound-matter interactions, we need to design scatterers that can provide strong bianisotropic coupling, and possibly reach the fundamental bound derived here. In the following, we show that it is possible to systematically design resonant particles that operate at the bound of maximum Willis coupling coefficient. Let us consider a 2D cylindrical particle with maze-like channels, as shown in Fig. 1a. If the particle is excited by a standing acoustic wave at a point where the applied velocity has its maximum, and the pressure is almost zero, we have vibration of air flow going in and out at the two channel outlets of the particle with volume velocities shown in Fig. 1b. The overall polarization of the particle due to the external excitation can be decomposed into dipole and monopole moments (see Fig. 1b). Let us now consider two special cases of this general scenario. Figure 1c shows the example of a symmetric case where $A = B$ (i.e., symmetric channel apertures). As it is seen from the figure, only the dipole moment will be excited in this scenario due to the external velocity field, indicating that the particle is non-bianisotropic. Interestingly, this is independent of the asymmetry of the internal maze of the particle, and it is guaranteed to first-order approximation as long as the apertures at the two sides are symmetric. On the other hand, figure 1d shows the case where the apertures are extremely asymmetric, with $B = 0$. For this scenario, the external velocity field can produce a dipole response and, in addition, a non-zero monopole response, indicating that the particle is bianisotropic, since the local velocity field induces a monopole response in addition to dipole. The subsets in Figs. 2a, 2b, and 2c show three different topologies for such a bianisotropic particle, which can provide different levels of bianisotropic coupling. The particle shown in the subset of Fig. 2a, is a symmetric particle whose bianisotropic response is weak. On the other hand, the particle shown in the subset of Fig. 2b is a bianisotropic particle, however its bianisotropic coupling cannot reach the bound introduced in the previous section, because the asymmetry is not strong. Finally, the particle shown in the subset of Fig. 2c, provides a very strong bianisotropic coupling that can reach the theoretical bound at its resonance frequencies, since one of the apertures is removed. Interestingly, this design reaches the bound at each resonance. The results in Fig. 2 indicates that different levels of bianisotropic coupling required for different applications can be realized using the proposed design. In Fig. 2d, we also show the case of a bianisotropic resonant particle with similar asymmetric response in both *x*- and *y*- directions. In this case, the general bound in Eq. (7) applies, where $\alpha_x^{vp\prime} = \alpha_y^{vp\prime}$. Indeed, also this particle reaches the bound for the sum of the two off-diagonal polarizability elements, showing that our proposed design approach for maximally bianisotropic scatterers works also in the case of multiple resonances along different axes.

The systematic design of optimal Willis scatterers introduced here enables translating many of the fascinating opportunities enabled by bianisotropy from electrodynamics to acoustics. In electromagnetic metamaterials, bianisotropic inclusions have been exploited to realize asymmetric absorbing metasurfaces, one-way transparent metasurfaces, and metagratings for perfect control of reflection and refraction [27]-[29]. Exploring analogous effect in acoustics, we can translate many of these applications to manipulate sound in unprecedented ways. Recently, in electrodynamics it was revealed that conventional gradient metasurfaces designed based on generalized laws of reflection and refraction [33], suffer from fundamental limits on conversion efficiency. A similar limitation has been found in the case of acoustic gradient metasurfaces [34]-[35]. In electrodynamics, metagratings have been proposed to address this issue. These structures are periodic arrays of properly designed bianisotropic inclusions that enable ultimate control over the reflected wave [29], with unitary efficiency even for anomalous reflection towards very large angles.

Here, we translate this idea into acoustics and design an acoustic metagrating based on our optimal Willis scatterers, capable of rerouting the reflected waves into extreme angles with unitary efficiency. Figure 3 shows the schematic of the designed metagrating, which is composed of a periodic array of the Willis scatterers, similar to Fig. 2, located with a distance from a hard plate. As an example, we design a metagrating to reroute a normally incident wave to $72°$. When a periodic array of bianisotropic inclusions over a hard plate is illuminated with a normally incident wave, different Floquet diffraction modes can scatter power away. By fixing the periodicity $b = \lambda/\sin\theta_{01}$ so as to align the 0(-1) order to the desired direction, we ensure that only three diffraction orders exist, the normal reflection, the desired reflection 0(-1) and the symmetric channel 01. We follow a design procedure similar to the one introduced in [29]. By choosing properly the distance of the bianisotropic particles from the hard plate, we can cancel the specular reflection represented by the 00 Floquet mode. As the next step, in order to have all the incident power channeled to the 0(-1) diffraction mode, the Willis scatterers are designed so as to radiate asymmetrically into 01 and 0(-1) modes when excited at normal incidence. As shown in Fig. 3, for a normally incident wave all sound is reflected into the desired direction (i.e., $\theta_{0(-1)} = -72°$), without any parasitic reflections.

In conclusion, we have introduced an analytical model for general acoustic scatterers. We studied the restrictions imposed by reciprocity and energy conservation on the polarizabilities of these particles. Using the proposed theory, we derived a theoretical bound for the acoustic bianisotropic coupling coefficient. It was proven that the bianisotropic coupling in acoustics can be of the same order as the diagonal components of the polarizability tensor. This finding can pave the way to translate many fascinating functionalities enabled by bianisotropy from electrodynamics to acoustics. Furthermore, we proposed realistic scatterers that can provide maximum bianisotropic coupling and reach the theoretical bound introduced in this Letter. As an application for the proposed inclusions, we employed them as building blocks of an acoustic metagrating that can reroute an illuminating wave to extreme directions in reflection with unitary efficiency.

This work was supported by the National Science Foundation.

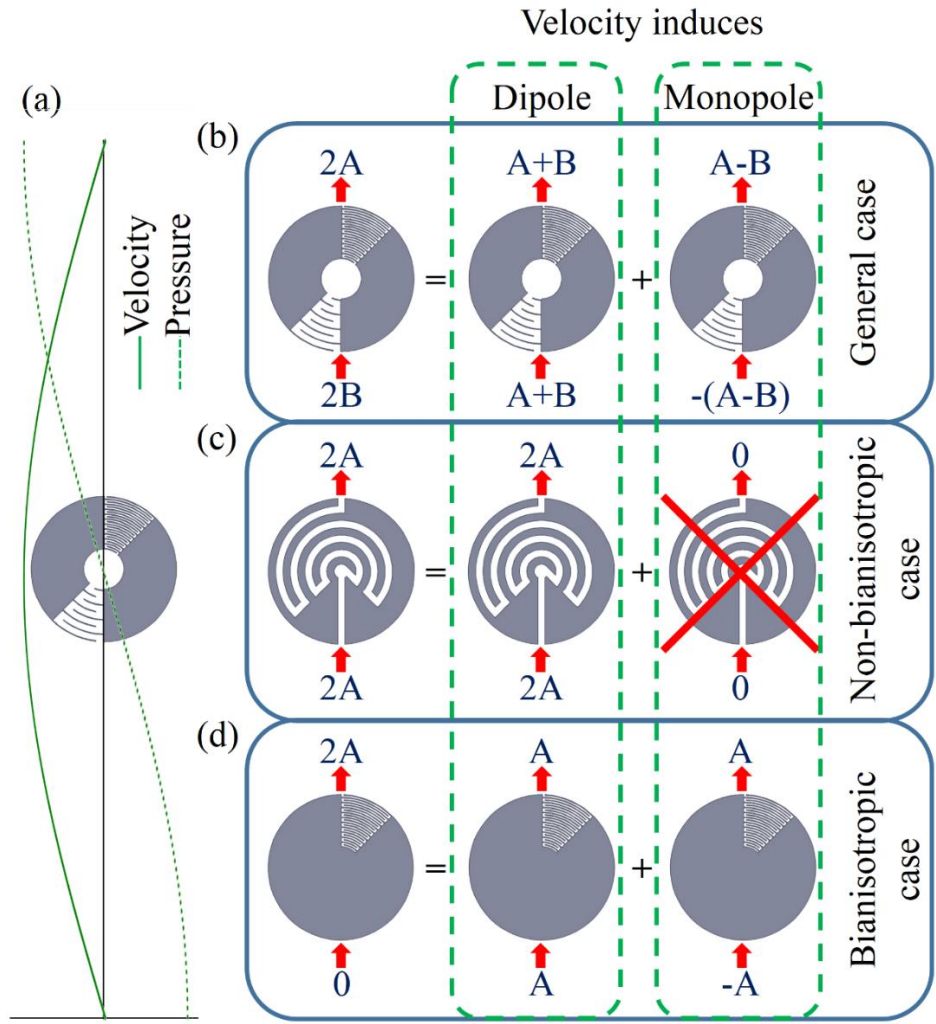

Fig. 1. (a) A general bianisotropic particle located in a standing acoustic wave. Physical interpretation of bianisotropy for a (b) general, (c) symmetric, and (d) asymmetric acoustic inclusions.

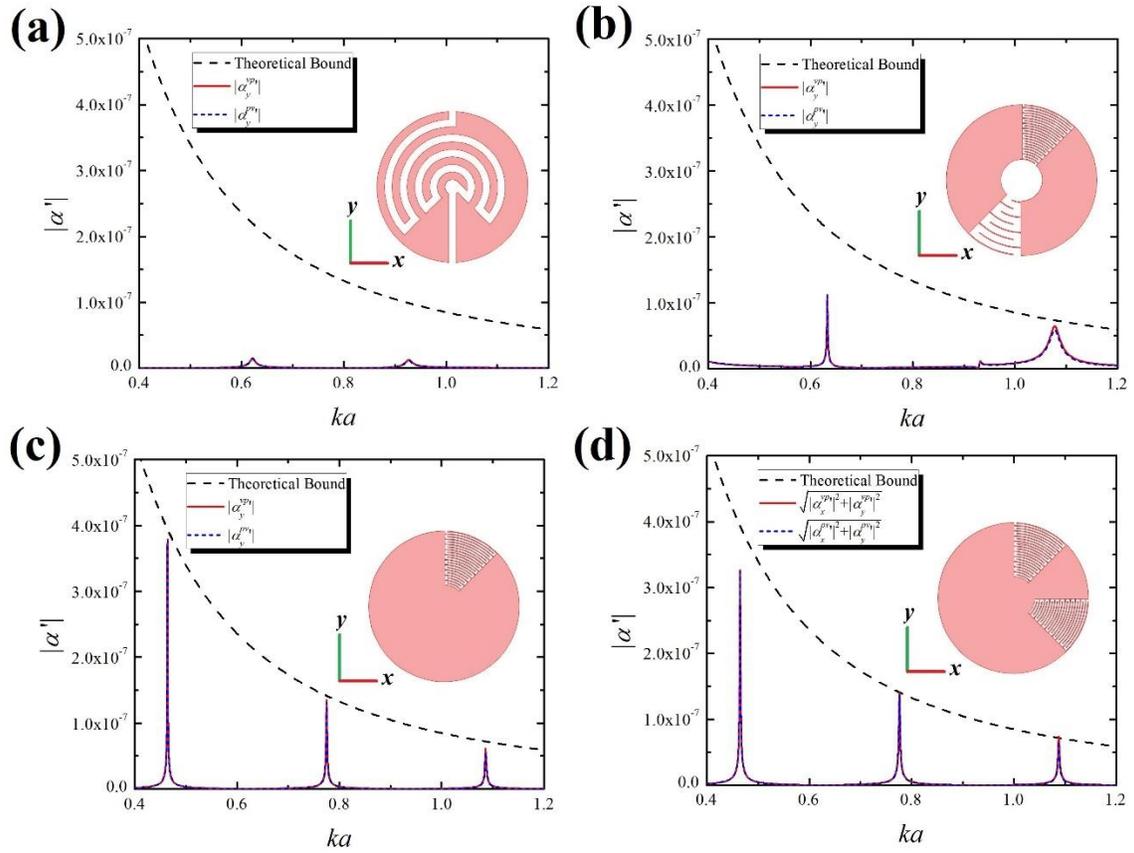

Fig. 2. (a), (b), (c) and (d) Comparison of cross-coupling terms for different biansiotropic inclusions providing different levels of bianisotropic coupling.

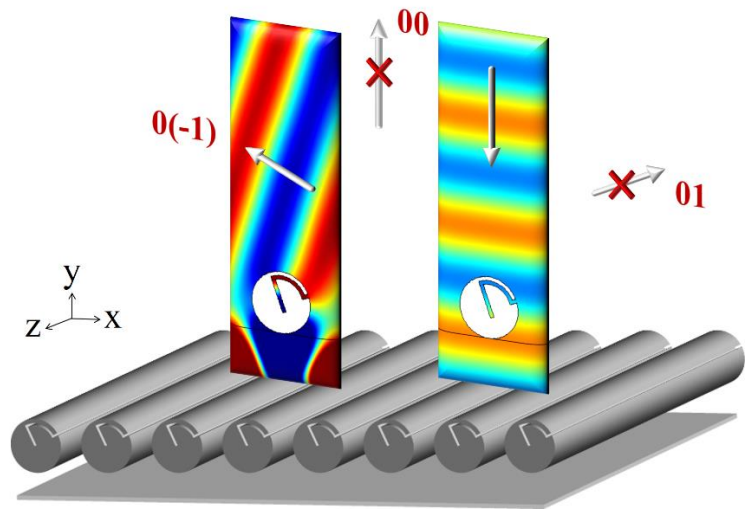

Fig. 3. Schematic of the designed acoustic metagrating, and simulated distribution of incident and reflected pressure fields.